\documentclass[smus]{snow2e}
\usepackage{graphics,epsfig,rotating,fleqn}

\begin{document}

\title{$t\bar{t}$ Production via Vector Boson Fusion at High Energy
$e^+e^-$ Colliders \thanks{This research
was supported in part by the Natural Sciences and Engineering
Research Council of Canada.}}

\author{Mikul\'{a}\v{s} Gintner and Stephen Godfrey \\
{\it Ottawa-Carleton Institute for Physics} \\
{\it Department of Physics, Carleton University, Ottawa CANADA, K1S 5B6} }

\maketitle

\thispagestyle{empty}\pagestyle{empty}

\begin{abstract}
We examine $t\bar{t}$ production via vector boson fusion at high
energy $e^+e^-$ colliders
using the effective vector-boson approximation.
We show cross sections as functions of CM energy for various
Higgs masses ranging from 100~GeV up to 1~TeV, and also for $M_H=\infty$
which corresponds to the LET. We give expressions for
$\sigma(V_i V_j \to t\bar{t})$
in the $2M_{W,Z}/\sqrt{s} = 0$ approximation and
show how this approximation effects the results.
\end{abstract}

\section{Introduction}
Understanding the mechanism responsible for electroweak symmetry
breaking is one of the most pressing problems in particle physics.
Quite generally, there are two scenarios.  In the first, the Higgs
boson is light and at high energies the weak sector remains weakly
coupled. In this scenario the Higgs boson should be observed at one of
the high energy colliders under consideration.
In the second scenario the Higgs boson is heavy and at high
energies the weak sector becomes strongly interacting.  If this were
the case a new spectroscopy is likely to manifest itself at energies
beyond a TeV.  In our view this is a very intriguing possibility.
Unfortunately,  it will be very difficult to detect and even more
difficult to understand.  Considerable work has been devoted to
understanding this problem using vector boson scattering at high
energy colliders with questionable success.
However, it has been shown that
$t\bar{t}$ production via vector boson fusion at high energy colliders
also provides a potentially powerful tool for understanding a strongly
interacting weak sector \cite{kauffman} and this subject has attracted
growing interest \cite{barklow,han}.
This is simply a manifestation
of the equivalence theorem where longitudinal bosons take on the
couplings of the scalars from which they acquire mass and a
consequence of
the fact that the Higgs boson couples most strongly to the most
massive particle available.

Due to the tremendous complexity of the algebra and expressions involved in
exact calculations of $t\bar{t}$ production via vector boson fusion,
the full analysis of these processes depends heavily on computer
techniques.  As a consequence, analytic expressions are not usually
presented.   Nevertheless, it is
often useful to have concise analytic expressions available for the purpose
of comparison and to gain insights into the process
under consideration.
An efficient way to simplify the complexity of the calculations
is to use the effective vector boson approximation (EVA). In this
approximation, vector bosons are treated as constituents of colliding
particles and the calculational requirements are reduced to finding
the cross section for the subprocess $V_iV_j \rightarrow t\bar{t}$,
with $V_i$  being a real particle, and convoluting this
cross section with the appropriate $W/Z$ distribution functions.
The EVA was used for the investigation of several processes in
the past, including heavy Higgs boson production, vector boson
pair scattering, and the production of heavy leptons at both
hadron and $e^+e^-$ colliders.
In all of these studies the EVA has been shown to be quite
accurate in the region of its expected validity
when compared to the full calculations.
Although some calculations have also been performed for
vector boson fusion $t\bar{t}$ production the picture is
far from complete. To the best of our knowledge
no full calculations of these processes exist in the literature.
The total cross section for
$e^+e^- \to \ell \bar{\ell} t\bar{t}$,
using the EVA, has been calculated by Kauffman at fixed CM energy
of 2~TeV \cite{kauffman}.
Closely related to this problem are calculations of heavy
fermion production at $pp$ colliders by Dawson and Willenbrock
\cite{dawson},
and heavy lepton production by Yuan ($pp$ and $e^+e^-$ colliders)
\cite{yuan},
and \'{E}boli {\it et al} ($pp$ collider) \cite{eboli}.
Approximate expressions for $V_iV_j \rightarrow t\bar{t}$
cross sections can be found in Kauffman's paper \cite{kauffman}
and can be adapted
from expressions for $V_iV_j \rightarrow L\bar{L}$ in
\'{E}boli {\it et al}'s paper \cite{eboli}.
Unfortunately, some of the published expressions contain errors.
The exact helicity amplitudes for
$V_iV_j \rightarrow L_1\bar{L_2}$ have been published in the
Appendix of Yuan's paper \cite{yuan}.

In this report we concentrate on $e^+e^-$ colliders and calculate
$e^+e^- \to \ell \bar{\ell} t\bar{t}$ cross sections as a
function of CM energy for various Higgs boson masses.
In our calculations we use the EVA with exact expressions
for $V_iV_j \rightarrow t\bar{t}$ cross sections.
For the $e^+e^- \to \ell \bar{\ell} t\bar{t}$
via $W_{L/T}W_{L/T}$ and $Z_{L/T}Z_{L/T}$ we evaluate
errors introduced by neglecting terms proportional
to $2M_{W,Z}/\sqrt{s}$ in $\sigma(WW/ZZ \rightarrow t\bar{t})$.
In the appendix we give expressions for 
$\sigma(V_iV_j \rightarrow t\bar{t})$
where we made the approximation  $2M_{W,Z}/\sqrt{s} = 0$
for $\sigma(WW \rightarrow t\bar{t})$ and
$\sigma(ZZ \rightarrow t\bar{t})$.

\section{Results}

\subsection{Cross Sections}

We calculated the full standard model expressions for the total
cross sections for:
\begin{itemize}
\item $W^+_{L/T}W^-_{L/T} \to t\bar{t}$
\item $Z_{L/T} Z_{L/T} \to t\bar{t}$
\item $\gamma Z_{L/T} \to t\bar{t}$
\item $\gamma \gamma \to t\bar{t}$
\end{itemize}
The expressions for these cross sections are given in the appendix
where $\sigma(WW \rightarrow t\bar{t})$ and
$\sigma(ZZ \rightarrow t\bar{t})$
are given in the approximation $2M_{W,Z}/\sqrt{s}=0$.
(The complete
expressions are prohibitively long to present in the limited space
allocated.)
In fig. 1 we show the cross sections for $e^+e^- \to \ell \bar{\ell}
t\bar{t}$ for the various subprocesses where the lepton $\ell$ is
either an $e$ or $\nu$ as appropriate.  To obtain these cross sections
we use the effective vector boson approximation (EVA),
take $m_t=175$~GeV, and take
$M_{t\bar{t}}>500$~GeV. The results are given for the full expressions
including all orders in $2M_W/\sqrt{s} $ and $2M_Z/\sqrt{s}$.
We will discuss the
effect of taking $M_{W,Z}/\sqrt{s} \to 0$ below.  Of these processes, it is
the subprocesses involving longitudinal vector bosons which are of
interest since the $V_L$ are, in some sense, the Higgs bosons.  The
subprocesses with transverse gauge bosons are backgrounds.  We also
note that the effective vector boson approximation is far from reliable
for transverse vector bosons and those results should be viewed
with caution.

\begin{figure}[t]
\leavevmode
\centerline{\epsfig{file=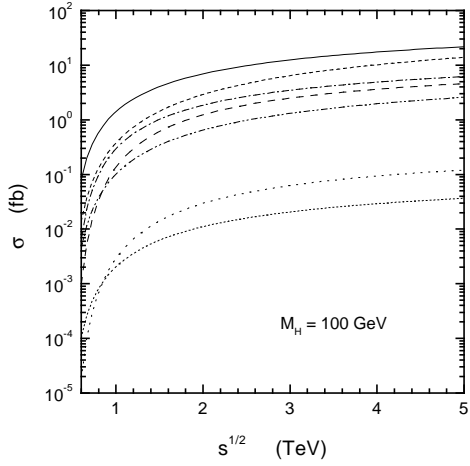,width=6.3cm,clip=}}
\caption{Cross sections for $t\bar{t}$ production via vector boson
fusion using the effective $W/Z/\gamma$ approximation with $M_H=100$~GeV.
The solid line is for $\gamma\gamma\to t\bar{t}$,
the dot-dashed line for $\gamma Z_L \to t\bar{t}$,
the dot-dot-dashed line for $\gamma Z_T\to t\bar{t}$,
the dashed line for $W_L W_L \to t\bar{t}$,
the short-dashed line for $W_T W_T \to t\bar{t}$,
the dotted line for $Z_L Z_L \to t\bar{t}$,
and the densely-dotted line for $Z_T Z_T \to t\bar{t}$}
\end{figure}

\subsection{Sensitivity to $M_H$}

Given that it is the processes $V_L V_L\to t\bar{t}$
and their sensitivity to the Higgs
boson mass that are of primary interest, in fig. 2 we show the cross
sections for $e^+ e^- \to \nu \bar{\nu} W_L W_L \to \nu \bar{\nu}t\bar{t}$
and $e^+ e^- \to e^+ e^- Z_L Z_L \to e^+ e^-  t \bar{t}$ for
$M_H=100$~GeV, $M_H=500$~GeV,
$M_H=1$~TeV, and $M_H=\infty$ (which corresponds to the
low energy theorem (LET)).
In fig. 3 we show the same
cross sections as a function of the
Higgs mass at fixed CM energy $\sqrt{s} = 1$~TeV.
For the $W_L W_L$ case
the cross sections at $\sqrt{s}=1$~TeV are  $\sim 0.1$~fb,
$\sim 0.4$~fb, and $\sim 0.3$~fb for $M_H=100$~GeV, 1~TeV, and $\infty$
respectively.  For the expected yearly integrated
luminosity of 200~fb$^{-1}$ these should be distinguisheable.
However, once $t-$quark detection efficiencies and
kinematic cuts to reduce backgrounds are included the situation is not
so clear.  We remind the reader that we already included a cut of
$M_{t\bar{t}}>500$~GeV and reducing this may increase the cross
section enough to distinguish the cases.  Although
in fig. 3 the cross section
is the same for certain values less than and greater than
$M_H\simeq 500$~GeV this does not concern us here since
if $M_H<500$~GeV the Higgs boson should be observed directly at the LHC.
As the centre of mass increases, the
cross section increases to 6~fb and 3~fb for $M_H=1$~TeV and
$M_H=\infty$ respectively at $\sqrt{s}=2$~TeV and $\sim$30~fb and
$\sim 15$~fb respectively at $\sqrt{s}=5$~TeV.  Thus,  the different
scenarios should be distinguisheable at these energies.

\begin{figure}[t]
\leavevmode
\centerline{\epsfig{file=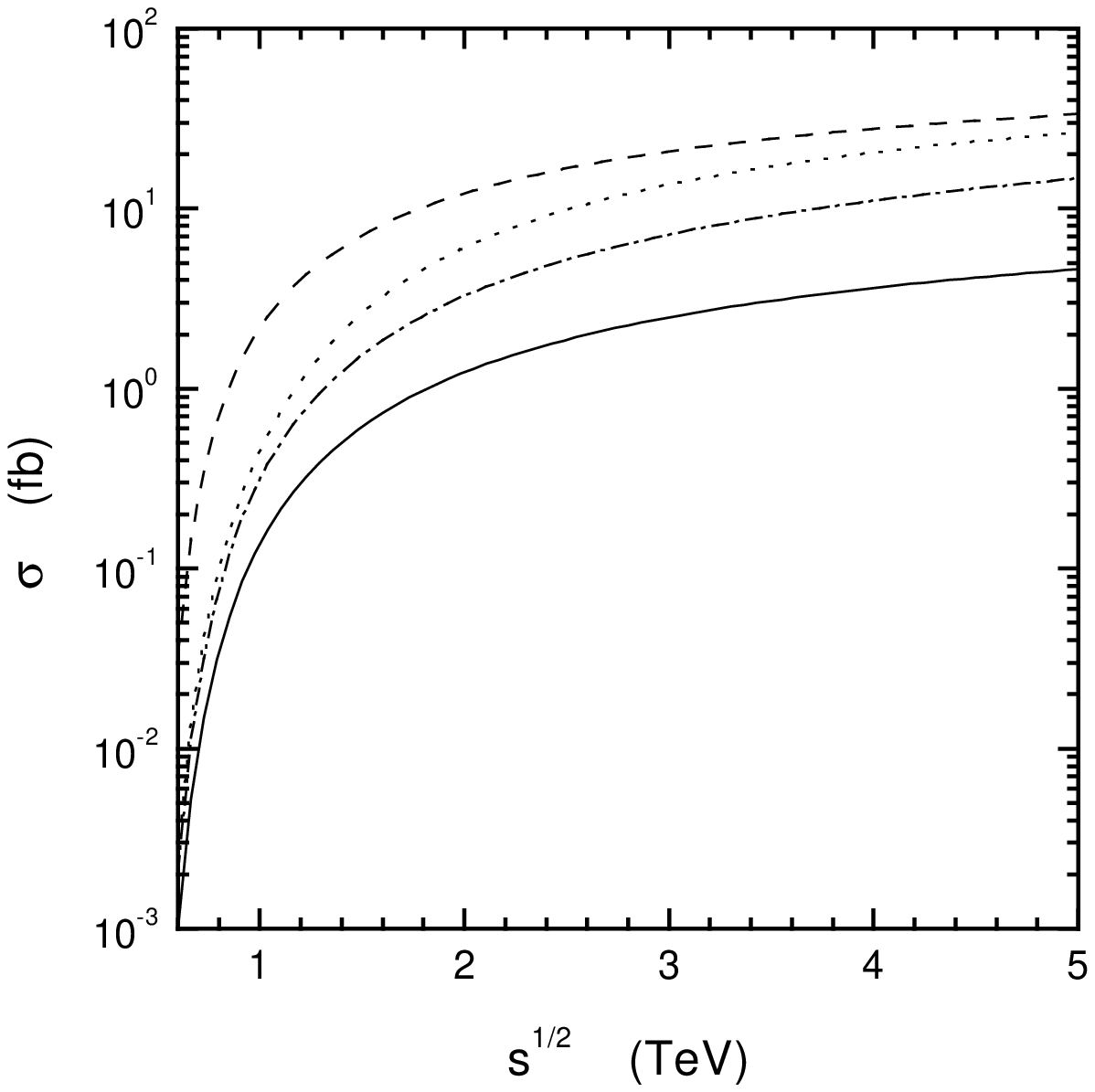,width=6.3cm,clip=}}
\centerline{\epsfig{file=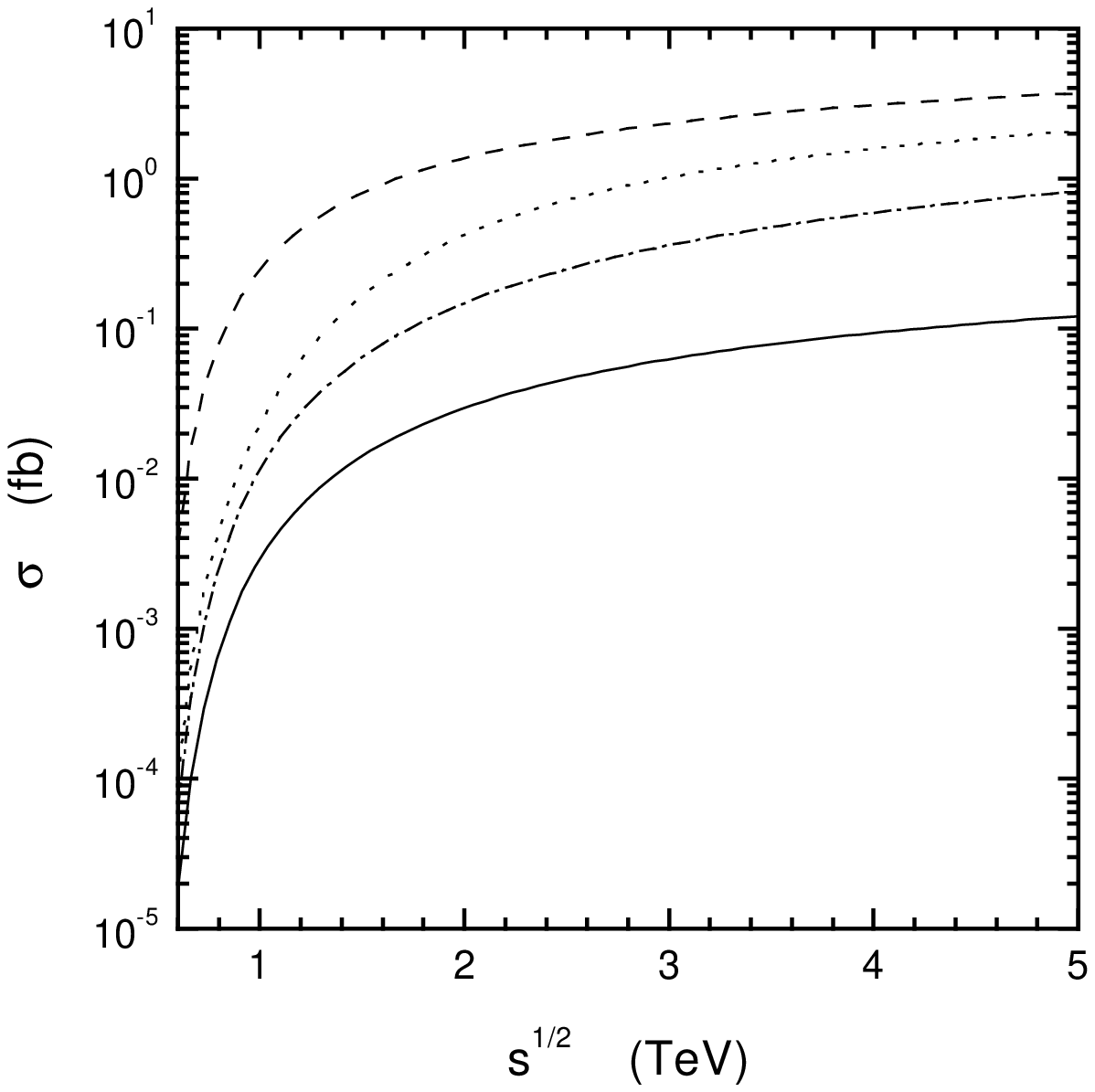,width=6.2cm,clip=}}
\caption{$\sigma(e^+ e^-\to \nu\bar{\nu} W_L W_L \to \nu\bar{\nu}t\bar{t})$
(upper figure) and
$\sigma(e^+ e^- \to e^+ e^- Z_L Z_L \to e^+ e^- t\bar{t})$ (lower figure)
vs $\sqrt{s}$ for several values of $M_H$. The solid lines
are for$M_H=100$~GeV, the dashed lines for $M_H=500$~GeV, the dotted
lines for $M_H=1$~TeV, and the dot-dashed lines for $M_H=\infty$ (LET).}
\end{figure}

\begin{figure}[h]
\leavevmode
\centerline{\epsfig{file=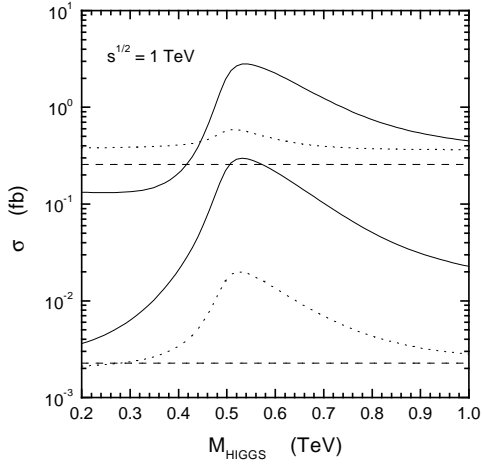,width=6.3cm,clip=}}
\caption{$\sigma(e^+ e^-\to \nu\bar{\nu} W_L W_L \to \nu\bar{\nu}t\bar{t})$
and $\sigma(e^+ e^- \to e^+ e^- Z_L Z_L \to e^+ e^- t\bar{t})$
for $\sqrt{s}=1$~TeV as a function of $M_H$.  The upper set of curves
is for $WW$ fusion and lower set of curves for $ZZ$ fusion.  
The solid curves are for $W_L W_L$ ($Z_L Z_L$) fusion, the
dashed curves for $W_L W_T$ ($Z_L Z_T$) fusion,  and the dotted curves
for $W_T W_T$ ($Z_T Z_T$) fusion.}
\end{figure}

\subsection{Errors Introduced with Approximations}

Although our figures are obtained using the complete expressions for
the subprocess cross sections, for brevity, we have given in the appendix
expressions in the limit that $2M_{W,Z}/\sqrt{s} =0$
for $WW \rightarrow t\bar{t}$ and $ZZ \rightarrow t\bar{t}$.
The complete expressions are not only lengthy but their
calculation is sufficiently complex
to diminish the simplifying motivation for using the EVA.
To help decide whether it is necessary to use the
full expressions
rather than the approximate ones shown in the appendix
and to know how large an error is introduced by
neglecting terms proportional to $2M_{W,Z}/\sqrt{s}$,
the differences between the exact and approximate expressions
for $e^+e^- \to \ell \bar{\ell} t\bar{t}$ via $WW$ and $ZZ$ fusion
are shown in fig. 4 as functions of the CM energy and the Higgs mass.
For CM energies between 0.6~TeV and 5~TeV and Higgs masses between 
100~GeV and $\infty$, 
deviations for $W_LW_L$ and $Z_L Z_L$ range up to about 10\%.
This is comparable
to the precision of the EVA which is claimed to be better than 10\%.
(Of course, the precision of EVA for
$e^+e^- \to \ell \bar{\ell} t\bar{t}$ in particular
has not been established yet.)
Errors for $W_TW_T$,  $W_LW_T$, $Z_TZ_T$, and $Z_LZ_T$ 
range up to 24\%, 13\%, 15\%, and 3\% respectively.  However,
as has been mentioned earlier, the EVA is expected to be less reliable
for the $LT$ and $TT$ modes than the $LL$ modes.
The errors tend to diminish  with energy but not dramaticaly
and not always uniformly. 
The convergence of the approximation is slowed down by the convolution 
of $\sigma(WW/ZZ\to t\bar{t})$ with the distribution functions which 
have large contributions from low energy $V_i V_j$ collisions where the 
approximate expressions for $\sigma(WW/ZZ \to t\bar{t})$ are not as 
accurate.

\begin{figure}[t]
\leavevmode
\centerline{\epsfig{file=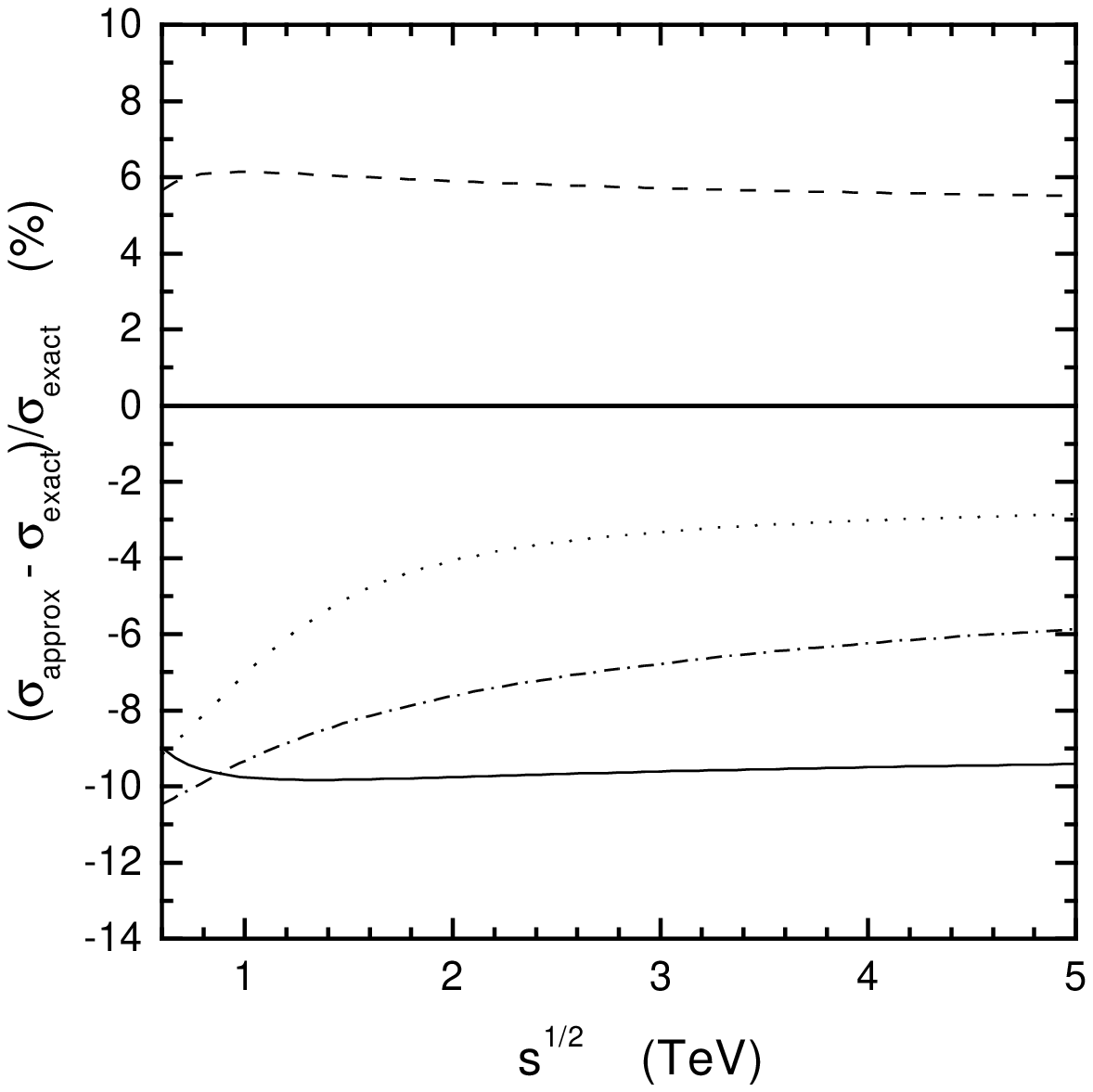,width=6.8cm,clip=}}
\centerline{\epsfig{file=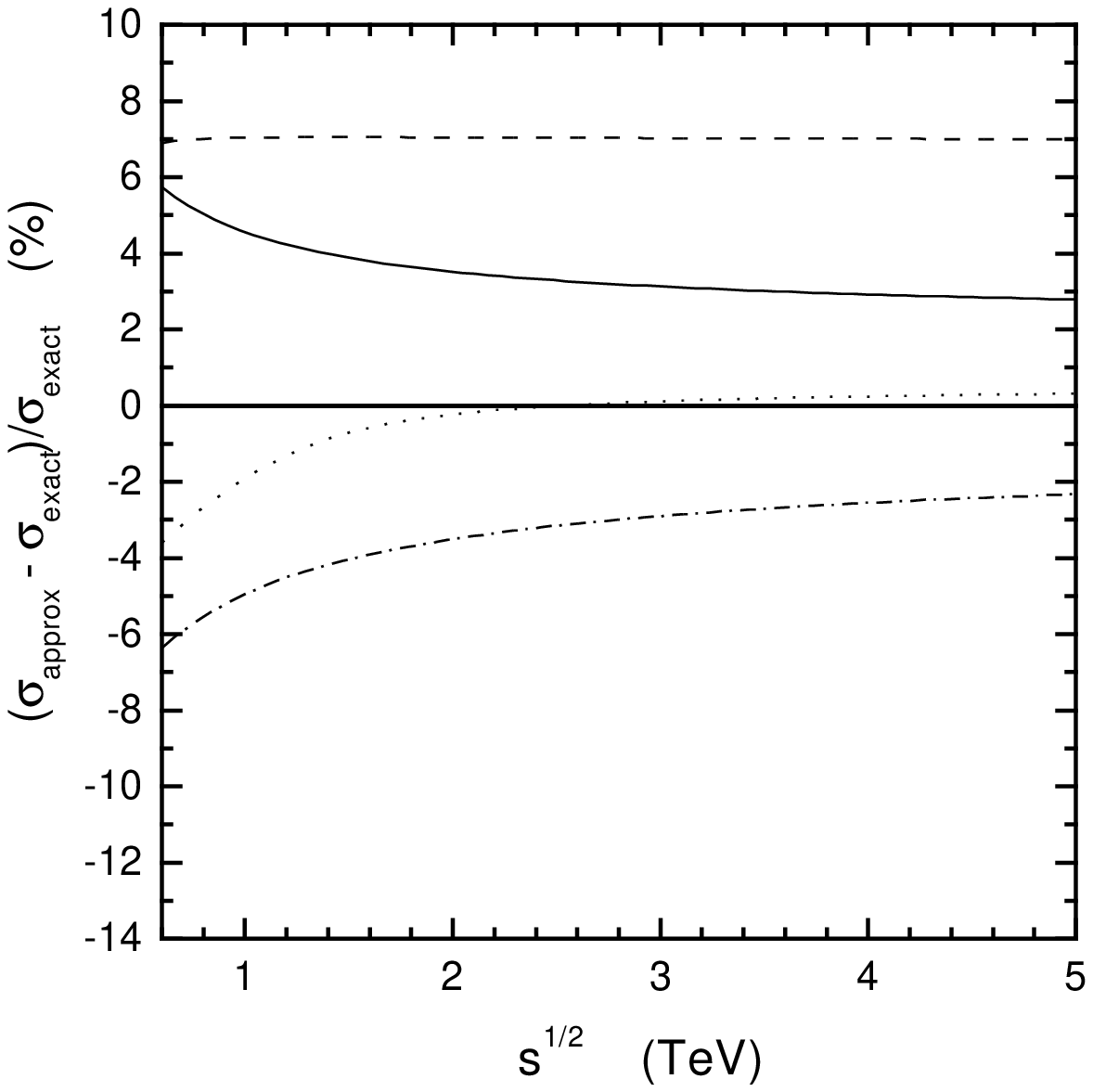,width=6.3cm,clip=}}
\caption{Errors introduced by taking $2 M_{W,Z}/\sqrt{s} \to 0$ for
the $W_L W_L $ mode (upper set of curves) and the $Z_L Z_L $ mode
(lower set of curves) for $e^+e^-\to \ell\bar{\ell} t\bar{t}$.  
In both cases the solid curves are
for $M_H=100$~GeV, the dashed curves for $M_H=500$~GeV, the dotted
curves for $M_H=1$~TeV, and the dot-dashed curves for $M_H=\infty$.}
\end{figure}

\section{Acknowledgements}

The authors thank Tim Barklow and Tao Han for helpful conversations and
communications.

\appendix
\section{Cross Section Formulas}

In this appendix we give the formulas for the
tree-level cross sections for
$V_iV_j\to t\bar{t}$.

\paragraph{\underline{$\gamma \gamma \rightarrow t \bar{t} $:}}

\begin{equation}
\sigma(\gamma \gamma \rightarrow t \bar{t} ) =
 \frac{4 \pi \alpha^2 Q_t^4 N_c}{s}
[(1+x_t^2-\frac{x_t^4}{2}) {\cal L} - \beta_t (1+x_t^2)]
\end{equation}

\paragraph{\underline{$\gamma Z_L \rightarrow t \bar{t} $:}}

\begin{eqnarray}
\lefteqn{\sigma(\gamma Z_L \rightarrow t \bar{t} ) =
\frac{4 \pi \alpha^2 Q_t^2 N_c} {s_W^2 c_W^2 s (1-M_Z^2/s)^3}
\frac{m_t^2}{M_Z^2}} \quad \nonumber \\
& & \times \{ [ a_t^2 ( 1-x_Z^2 + \frac{x_t^2 x_Z^2}{4} ) -
(a_t^2+2 v_t^2) \frac{x_t^2 x_Z^2}{16} \frac{M_Z^2}{m_t^2} ] {\cal L}
\nonumber \\
& & + \frac{1}{4} \beta_t x_Z^2
[ 2 a_t^2 + (a_t^2+v_t^2) \frac{M_Z^2}{m_t^2} ] \}
\end{eqnarray}

\paragraph{\underline{$\gamma Z_T \rightarrow t \bar{t} $:}}

\begin{eqnarray}
\lefteqn{\sigma(\gamma Z_T \rightarrow t \bar{t} ) =
\frac{\pi \alpha^2 Q_t^2 N_c} {s_W^2 c_W^2 s (1-M_Z^2/s)^3} }
\quad\nonumber \\
& & \times \{ (a_t^2+v_t^2)
[ ({\cal L}-\beta_t) (1+\frac{1}{16} x_Z^4) -  \frac{1}{2} \beta_t x_Z^2]
\nonumber \\
& & + (a_t^2-v_t^2) [ \beta_t - (1-\frac{1}{2} x_t^2) {\cal L}] x_t^2 \}
\end{eqnarray}

\paragraph{\underline{$W_L W_L \rightarrow t \bar{t} $:}}
($x_W =0$ approximation)

\begin{equation}
\sigma(W_L W_L \rightarrow t \bar{t}) = \sum_{i} \sigma_i^{LL}
\quad \mbox{where}
\end{equation}


\begin{equation}
\sigma_{\gamma \gamma}^{LL}  = \frac{\pi \alpha^2 Q_t^2 N_c}{3}
\frac{s}{M_W^4} \beta_t (1+\frac{1}{2} x_t^2)
\end{equation}

\begin{eqnarray}
\sigma_{ZZ}^{LL} & = & \frac{\pi \alpha^2 N_c} {12 s_W^4 (1-M_Z^2/s)^2}
\frac{s}{M_W^4}  \beta_t \nonumber \\
& & \qquad \times [v_t^2 (1+\frac{1}{2} x_t^2) + a_t^2 (1-x_t^2)]
\end{eqnarray}

\begin{equation}
\sigma_{HH}^{LL}  = \frac{\pi \alpha^2 N_c} {8 s_W^4} \frac{m_t^2}{M_W^4}
\beta_t^3 s^2 \chi_H
\end{equation}

\begin{equation}
\sigma_{bb}^{LL}  = \frac{\pi \alpha^2 N_c} {24 s_W^4}
\frac{s}{M_W^4} \beta_t^2 [\beta_t (1+\frac{3}{2} x_t^2) +
\frac{3}{4} x_t^4 {\cal L} ]
\end{equation}

\begin{equation}
\sigma_{\gamma Z}^{LL}  = \frac{\pi \alpha^2 Q_t N_c v_t}
{3 s_W^2 (1-M_Z^2/s)} \frac{s}{M_W^4} \beta_t (1+\frac{1}{2} x_t^2)
\end{equation}

\begin{equation}
\sigma_{Hb}^{LL}  =
- \frac{\pi \alpha^2 N_c} {4 s_W^4} \frac{m_t^2}{M_W^4}
[ \beta_t (1- \frac{x_t^2}{2} ) -
\frac{x_t^4}{4} {\cal L} ] s (s-M_H^2) \chi_H
\end{equation}

\begin{equation}
\sigma_{\gamma b}^{LL}  =
- \frac{\pi \alpha^2 Q_t N_c} {6 s_W^2} \frac{s}{M_W^4}
[ \beta_t (1- \frac{x_t^2}{4} - \frac{3}{8} x_t^4 ) -
\frac{3}{16} x_t^6 {\cal L} ]
\end{equation}

\begin{eqnarray}
\sigma_{Zb}^{LL} & = & -
\frac{\pi \alpha^2 N_c (a_t+v_t)}{12 s_W^4 (1-M_Z^2/s)} \frac{s}{M_W^4}
\nonumber \\
& & \qquad \times [ \beta_t (1- \frac{1}{4} x_t^2 - \frac{3}{8} x_t^4 ) -
\frac{3}{16} x_t^6 {\cal L} ]
\end{eqnarray}

\begin{equation}
\sigma_{\gamma H}^{LL} = \sigma_{ZH}^{LL} = 0
\end{equation}

\paragraph{\underline{$W_T W_T \rightarrow t \bar{t} $:}}
($x_W =0$ approximation)

\begin{equation}
\sigma(W_T W_T \rightarrow t \bar{t}) = \sum_{i} \sigma_i^{TT}
\qquad \mbox{where}
\end{equation}


\begin{equation}
\sigma_{\gamma \gamma}^{TT} =
\frac{2 \pi \alpha^2 Q_t^2 N_c}{3s} \beta_t (1+\frac{1}{2} x_t^2)
\end{equation}

\begin{eqnarray}
\sigma_{ZZ}^{TT} & = &
\frac{\pi \alpha^2 N_c}
{6 s_W^4 s (1-M_Z^2/s)^2} \beta_t \nonumber \\
& & \qquad \times [v_t^2 (1+\frac{1}{2} x_t^2) + a_t^2 (1-x_t^2)]
\end{eqnarray}

\begin{equation}
\sigma_{HH}^{TT}  = \frac{\pi \alpha^2 N_c} {4 s_W^4}
m_t^2 \beta_t^3 \chi_H
\end{equation}

\begin{eqnarray}
\sigma_{bb}^{TT} & = & \frac{\pi \alpha^2 N_c}{2 s_W^4 s}
[\beta_t (-\frac{7}{6}+\frac{17}{12}x_t^2-\frac{1}{2}x_t^4) 
\nonumber\\
& & \qquad \qquad
+ (1-x_t^2+\frac{7}{8}x_t^4-\frac{1}{4}x_t^6) {\cal L} ]
\end{eqnarray}

\begin{equation}
\sigma_{\gamma Z}^{TT}  = \frac{2 \pi \alpha^2 Q_t N_c v_t}
{3 s_W^2 s (1-M_Z^2/s)} \beta_t (1+\frac{1}{2} x_t^2)
\end{equation}

\begin{equation}
\sigma_{Hb}^{TT} = \frac{\pi \alpha^2 N_c}{s_W^4}\frac{m_t^4}{s^2}
[  (1-\frac{x_t^2}{2}) {\cal L} - \beta_t ]
(s-M_H^2) \chi_H
\end{equation}

\begin{eqnarray}
\sigma_{\gamma b}^{TT} & = & - \frac{\pi \alpha^2 Q_t N_c}{3 s_W^2 s}
[ \beta_t (1+\frac{5}{4}x_t^2-\frac{3}{8}x_t^4)
\nonumber \\
& & \qquad \qquad + (\frac{3}{4}-\frac{3}{16}x_t^2)x_t^4 {\cal L} ]
\end{eqnarray}

\begin{eqnarray}
\sigma_{Zb}^{TT} & = &
- \frac{\pi \alpha^2 N_c}{6 s_W^4 s (1-M_Z^2/s)} \nonumber \\
& & \qquad \times \{ \beta_t [ a_t (1-\frac{1}{4}x_t^2-\frac{3}{8}x_t^4)+
\nonumber \\
& & \qquad\qquad  v_t (1+\frac{5}{4}x_t^2-\frac{3}{8}x_t^4) ]
\nonumber \\
& & \qquad + \frac{3}{4}x_t^4
[ v_t (1-\frac{1}{4}x_t^2)-\frac{1}{4}a_tx_t^2] {\cal L} \}
\end{eqnarray}

\begin{equation}
\sigma_{\gamma H}^{TT} = \sigma_{ZH}^{TT} = 0
\end{equation}

\paragraph{\underline{$W_L W_T \rightarrow t \bar{t} $:}}
($x_W =0$ approximation)

\begin{equation}
\sigma(W_L W_T \rightarrow t \bar{t})  = \sum_{i} \sigma_i^{LT}
\qquad \mbox{where}
\end{equation}
\begin{equation}
\sigma_{\gamma \gamma}^{LT}  = \frac{4 \pi \alpha^2 Q_t^2 N_c}{3 M_W^2}
\beta_t (1+\frac{1}{2} x_t^2)
\end{equation}

\begin{eqnarray}
\sigma_{ZZ}^{LT} & = &
\frac{\pi \alpha^2 N_c}{3 s_W^4 (1-M_Z^2/s)^2 M_W^2} \beta_t
\nonumber\\
& & \qquad \times [v_t^2 (1+\frac{1}{2} x_t^2) + a_t^2 (1-x_t^2)]
\end{eqnarray}

\begin{eqnarray}
\sigma_{bb}^{LT} & = & \frac{\pi \alpha^2 N_c}{6 s_W^4 M_W^2}
[\beta_t (1-x_t^2+\frac{3}{4} x_t^4)
\nonumber\\
& & \qquad \qquad  + \frac{3}{4} x_t^2(1-x_t^2+\frac{1}{2} x_t^4) {\cal L} ]
\end{eqnarray}

\begin{equation}
\sigma_{\gamma Z}^{LT}  = \frac{4 \pi \alpha^2 Q_t N_c v_t}
{3 s_W^2 (1-M_Z^2/s) M_W^2}\beta_t (1+\frac{1}{2} x_t^2)
\end{equation}

\begin{equation}
\sigma_{\gamma b}^{LT} = \frac{2 \pi \alpha^2 Q_t N_c}{3 s_W^2 M_W^2}
[ \beta_t (1+ \frac{x_t^2}{8} + \frac{3}{16} x_t^4 ) +
\frac{3}{32} x_t^6 {\cal L} ]
\end{equation}

\begin{eqnarray}
\sigma_{Zb}^{LT} & = & - \frac{\pi \alpha^2 N_c}{3 s_W^4 (1-M_Z^2/s) M_W^2}
\nonumber \\
& & \qquad \times \{ \beta_t
[ a_t (1- \frac{5}{8} x_t^2 + \frac{3}{16} x_t^4 ) + \nonumber\\
& & \qquad\qquad  v_t (1+ \frac{1}{8} x_t^2 + \frac{3}{16} x_t^4 )]
\nonumber \\
& & \qquad \qquad - \frac{3}{8} x_t^4
[a_t (1-\frac{1}{4}x_t^2)-\frac{1}{4}v_tx_t^2] {\cal L} \}
\end{eqnarray}
\begin{eqnarray}
\sigma_{HH}^{LT} = \sigma_{Hb}^{LT} =
\sigma_{\gamma H}^{LT} = \sigma_{ZH}^{LT} = 0
\end{eqnarray}

\paragraph{\underline{$Z_LZ_L \rightarrow t \bar{t} $:}}
($x_Z =0$ approximation)
\begin{eqnarray}
\sigma(Z_L Z_L \rightarrow t \bar{t}) & = & \sum_{i} \sigma_i^{LL}
\qquad \mbox{where}
\end{eqnarray}
%
\begin{eqnarray}
\sigma_{ss}^{LL} & = & \frac{\pi \alpha^2 N_c}{8 s_W^4 c_W^4}
\frac{m_t^2}{M_Z^4} \beta_t^3 s^2 \chi_H
\end{eqnarray}
\begin{eqnarray}
\sigma_{tt}^{LL} = \sigma_{uu}^{LL} & = &
\frac{\pi \alpha^2 N_c}{48 s_W^4 c_W^4}\frac{s}{M_Z^4} \beta_t
[(a_t^4+v_t^4)(1+\frac{1}{2}x_t^2) \nonumber\\
& & \qquad\qquad
+ 6a_t^2v_t^2(1-\frac{3}{2}x_t^2)]
\end{eqnarray}
\begin{equation}
\sigma_{st}^{LL} =
\sigma_{su}^{LL}  =
\frac{\pi \alpha^2 N_c a_t^2}
{2 s_W^4 c_W^4}
\frac{m_t^2}{M_Z^4}
(\beta_t-\frac{x_t^2}{2}{\cal L})s(s-M_H^2)\chi_H
\end{equation}
\begin{eqnarray}
\sigma_{tu}^{LL} & = & -\frac{\pi \alpha^2 N_c}{24 s_W^4 c_W^4}
\frac{s}{M_Z^4}
\nonumber \\
& & \qquad \times
\{ \beta_t
[a_t^4(1-\frac{23}{2}x_t^2)+v_t^4(1+\frac{1}{2}x_t^2) \nonumber \\
& & \qquad\qquad + 6a_t^2v_t^2(1-\frac{3}{2}x_t^2)] + 6a_t^4x_t^4{\cal L} \}
\end{eqnarray}
\paragraph{\underline{$Z_TZ_T \rightarrow t \bar{t} $:}}
($x_Z =0$ approximation)
\begin{equation}
\sigma(Z_T Z_T \rightarrow t \bar{t})  = \sum_{i} \sigma_i^{TT}
\qquad \mbox{where}
\end{equation}
%
%
\begin{eqnarray}
\sigma_{ss}^{TT} & = & \frac{\pi \alpha^2 N_c}{4 s_W^4 c_W^4} m_t^2
\beta_t^3 \chi_H
\end{eqnarray}
\begin{eqnarray}
\sigma_{tt}^{TT} & = & \sigma_{uu}^{TT}  =
\frac{\pi \alpha^2 N_c}{8 s_W^4 c_W^4 s} \nonumber\\
& & \hspace{-2mm} \times
\{ -\frac{4}{3}\beta_t [a_t^4(1-x_t^2)+v_t^4(1+2x_t^2)+12a_t^2v_t^2]
\nonumber \\
& & +[ a_t^4 (1-\frac{1}{2}x_t^2)+v_t^4(1+\frac{3}{2}x_t^2)
\nonumber \\
& & \qquad \qquad + 6a_t^2v_t^2(1+\frac{1}{2}x_t^2)] {\cal L} \}
\end{eqnarray}
\begin{equation}
\sigma_{st}^{TT} = \sigma_{su}^{TT}  =
\frac{\pi \alpha^2 N_c}{4 s_W^4 c_W^4} (v_t^2-a_t^2)\frac{m_t^2}{s}
\beta_t^2 {\cal L}(s-M_H^2)\chi_H
\end{equation}
\begin{eqnarray}
\sigma_{tu}^{TT} & = & \frac{\pi \alpha^2 N_c}{12 s_W^4 c_W^4 s}
\{ \beta_t [(a_t^4+v_t^4)(1+5x_t^2) \nonumber\\
& & \qquad\qquad + 6a_t^2v_t^2(1+x_t^2)] \nonumber \\
& & \qquad -\frac{3}{2}x_t^2 [(a_t^4+v_t^4)(1+x_t^2) \nonumber\\
& & \qquad\qquad + 6a_t^2v_t^2(1-\frac{1}{3}x_t^2)]{\cal L} \}
\end{eqnarray}
\paragraph{\underline{$Z_LZ_T \rightarrow t \bar{t} $:}}
($x_Z =0$ approximation)
\begin{equation}
\sigma(Z_L Z_T \rightarrow t \bar{t})  = \sum_{i} \sigma_i^{LT}
\qquad \mbox{where}
\end{equation}
%
%
\begin{equation}
\sigma_{ss}^{LT} = \sigma_{st}^{LT} = \sigma_{su}^{LT} = 0
\end{equation}
\begin{eqnarray}
\sigma_{tt}^{LT} =
\sigma_{uu}^{LT} & = &
\frac{\pi \alpha^2 N_c}{12 s_W^4 c_W^4 M_Z^2} \nonumber \\
& & \times \{ \beta_t [a_t^4(1-x_t^2)+v_t^4(1+\frac{1}{2}x_t^2)
\nonumber\\
& &  + 6a_t^2v_t^2(1+\frac{1}{4}x_t^2)]
 -3a_t^2v_t^2x_t^2 {\cal L} \}
\end{eqnarray}
\begin{eqnarray}
\sigma_{tu}^{LT} & = & -\frac{\pi \alpha^2 N_c}{6 s_W^4 c_W^4 M_Z^2}
\{ \beta_t [a_t^4(1+2x_t^2)
\nonumber \\
& &  +v_t^4(1+\frac{1}{2}x_t^2)+ 6a_t^2v_t^2(1+\frac{1}{4}x_t^2)]
\nonumber \\
& & \qquad
 -\frac{3}{2}x_t^2(a_t^4+3a_t^2v_t^2) {\cal L} \}
\end{eqnarray}
In all the above equations we used the substitutions
\begin{equation}
{\cal L}  =  \ln \left( \frac{1+\beta_t}{1-\beta_t} \right),
\quad
\chi_H  = \frac{1}{(s-M_H^2)^2+\Gamma_H^2 M_H^2}
\end{equation}
$\beta_{t,z,w} = \sqrt{1-x_{t,z,w}^2}$,
$x_t  =  2 m_t/\sqrt{s}$, $x_Z  = 2 M_Z/\sqrt{s}$, $x_W = 2 M_W/\sqrt{s}$,
$s_W  = \sin \theta_W$, $c_W = \cos \theta_W$,
$a_t = \frac{1}{2}$, $v_t = \frac{1}{2} - \frac{4}{3} s_W^2$,
$Q_t$ is the electric charge of the top quark in terms of $|e|$,
and $N_c$ is the number of colours.

%

\end{document}